\documentclass[10pt,conference,twocolumn]{IEEEtran}
\IEEEoverridecommandlockouts

\usepackage{cite}
\usepackage{amsmath,amssymb,amsfonts}
\usepackage{algorithm}
\usepackage{algpseudocode} 
\usepackage{graphicx}
\usepackage{textcomp}
\usepackage{xcolor}
\def\BibTeX{{\rm B\kern-.05em{\sc i\kern-.025em b}\kern-.08em
    T\kern-.1667em\lower.7ex\hbox{E}\kern-.125emX}}
\begin{document}
	
	\title{Enhancing Secret Key Generation for UAV Communications via Codeword Reconstruction\\
	\thanks{This work was supported in part by the Key Research and Development Plan of Shaanxi Province under Grant No. 2025CG-GJHX-06, the Fundamental Research Funds for the Central Universities and the National Key Laboratory of Wireless Communications Foundation under Grant IFN202501.}
	}

	\author{\IEEEauthorblockN{Yizhuo Wang\IEEEauthorrefmark{1}\IEEEauthorrefmark{5}, Qinghe Du\IEEEauthorrefmark{1}\IEEEauthorrefmark{2}\IEEEauthorrefmark{5}, Ning Shen\IEEEauthorrefmark{1}\IEEEauthorrefmark{5}, Xiao Tang\IEEEauthorrefmark{1}\IEEEauthorrefmark{5}, Shijiao Zhang\IEEEauthorrefmark{1}\IEEEauthorrefmark{5}, Lei Zhao\IEEEauthorrefmark{3}, Yang Hu\IEEEauthorrefmark{1}\IEEEauthorrefmark{4}}
		\IEEEauthorblockA{\IEEEauthorrefmark{1}\textit{School of Information and Communications Engineering, Xi’an Jiaotong University, Xi’an 710049, China}}
		\IEEEauthorblockA{\IEEEauthorrefmark{2}\textit{National Key Laboratory of Wireless Communications, Chengdu 611731, China}}
		\IEEEauthorblockA{\IEEEauthorrefmark{3}\textit{Xi’an Tianyi Intelligent Aircraft Technology Group Co.,LTD, Xi’an 710049, China}}
		\IEEEauthorblockA{\IEEEauthorrefmark{4}\textit{Shaanxi Vocational and Technical College, Xi’an 710049, China}}
		\IEEEauthorblockA{\IEEEauthorrefmark{5}\textit{Shaanxi Smart Networks and Ubiquitous Access Research Center, Xi’an 710049, China}\\
			Emails: \textit{Wyz0219@stu.xjtu.edu.cn, duqinghe@mail.xjtu.edu.cn, sn872416614@stu.xjtu.edu.cn,} \\ \textit{tangxiao@xjtu.edu.cn, zhshijiao@xjtu.edu.cn, zhaolei@tianyiair.com, sdlihe.lilium@163.com}}
	}
	\maketitle

\begin{abstract}
\par With the rapid advancement of unmanned aerial vehicle (UAV), ensuring the security of communication links among UAVs has become crucial. In this paper, we propose a novel physical layer key generation scheme based on channel codeword reconstruction. In UAV communications, the high mobility of aerial nodes leads to short channel coherence time, which together with noise causes inevitable channel estimation errors. These errors significantly degrades the performance of wireless channel–based key generation. Therefore, we propose a codeword construction algorithm that achieves a polarization characteristic, which effectively segregates reliable keys from unreliable ones. Compared to the existing quantization-based key generation scheme, our approach maximize the utilization of raw channel information and employ soft-decision decoding to generate key. Simulation results demonstrate that the proposed scheme reduces the key disagreement rate for legitimate users and increases the number of consistently generated keys. Furthermore, our method ensures a lower key consistency rate for eavesdropper, which guarantees system security.
\end{abstract}


\section{Introduction}
\par In recent years, advances in wireless communications have enabled the widespread deployment of unmanned aerial vehicle (UAV) in applications such as parcel delivery and logistics, emergency services, and environmental monitoring \cite{yao2021resource}. Benefiting from its high-mobility characteristics, UAV can also serve as a powerful enabler of emerging wireless technologies, such as integrated sensing and communications\cite{zheng2025optimal}. Moreover, to support collaborative perception among multiple UAVs, these systems often rely on wireless links for information exchange. However, such wireless links are inherently vulnerable to interference, spoofing, and eavesdropping, which may cause mission deviations or loss of control and, lead to casualties in severe cases \cite{ceviz2024survey}. Therefore, ensuring the security of UAV communications is of paramount importance.
\par As network scale and topological complexity increase, conventional cryptographic schemes introduce substantial overhead for key management and distribution \cite{sen2023cryptography}. This challenge is further compounded in UAV communications where nodes frequently join or leave, imposing stringent requirements on the key management. In contrast, physical-layer security techniques are an effective approach for secure UAV communications and have demonstrated significant success in both key generation \cite{sun2018review,sun2017physical} and opportunistic data transmission \cite{xiao2023secure}. Specifically, physical-layer key generation (PLKG) exploits channel reciprocity to directly generate secret keys over the air interface, which eliminates the need for complex key distribution protocols. Compared with ensuring security by exploiting the opportunistic advantage of time-varying channel quality, PLKG can be more smoothly integrated into existing cryptographic framework and it imposes less stringent requirements on channel quality, thus facilitating easier deployment. The inherent efficiency of PLKG makes it an ideal approach for UAV secure communications.
\par Despite the advantages of PLKG in UAV networks, its practical implementation faces significant challenges. PLKG necessitates frequent interactions between transmitter and receiver, leading to considerable latency, which is not suited for time-sensitive or information freshness-critical applications \cite{xiao2024statistical,xiao2023adaptive}. In addition, the high mobility inherent to UAV results in a short channel coherence time. This exacerbates channel estimation errors and leads to discrepancies between the uplink and downlink channel states information (CSI), consequently degrading reciprocity\cite{gao2024ris}. Furthermore, when the channel samples at both ends fall close to the quantization thresholds, small mismatches can lead to large quantization errors. This requires repeated information exchanges during key reconciliation, significantly reducing security \cite{li2019hybrid}. To overcome these challenges, the authors in \cite{liu2019secret} exploit the eigenvalues of channel covariance matrix to generate secret key. However, this scheme requires a specially antenna array. Torshizi et al. leverage the reciprocity of scattering matrix parameters and use MUSIC-based direction-of-arriva (DoA) estimation to extract feature \cite{torshizi2024exploiting}. In addition, reference \cite{he2022deep} proposes a key generation algorithm based on an adversarial autoencoder, which maps the channel samples into a reciprocal feature space by neural network. 
\par The above schemes either heavily rely on prior channel information or require specially hardware, which is not suitable for UAV communications. When directly applied, these methods will result in an increased Key Disagreement Rate (KDR) and a decreased Key Generation Rate (KGR). In this paper, we propose a PLKG scheme based on channel codeword reconstruction to address the existing degradation of channel reciprocity. The proposed scheme integrates the quantization and reconciliation to preserve the maximum amount of information in the channel samples. It arranges the random channel samples into a codeword-structured sequence and employs error-correcting code (ECC) to compensate for the effects of channel non-reciprocity. The main contributions of this work are summarized as follows.

\begin{itemize}
	\item To the best of our knowledge, this is the first work to reconstruct a channel with a codeword structure and employ soft-decision decoding to generate the key. Highly consistent key generation is achieved even in UAV communications with degraded channel reciprocity.
	\item We propose a novel permutation-table construction algorithm to establish a specific codeword structure at the receiver. Crucially, this algorithm achieves a polarization effect that separates reliable and unreliable bits, providing a solid basis for highly consistent key generation through tail truncation.
	\item Simulation results demonstrate that the proposed scheme achieves a significantly lower KDR and a higher KGR, owing to the superior error mitigation provided by the ECC. we evaluate the security performance of the scheme by considering the eavesdropper.
\end{itemize}

\section{System Model}

\par As shown in Fig.~1, our system consists of three nodes: a moving UAV node (Alice) operating as the transmitter, a legitimate user (Bob) and an eavesdropper (Eve). Alice and Bob share a bidirectional wireless link and communicate over the same frequency band utilizing Time Division Duplex (TDD) mode, while Eve passively eavesdrops and does not transmit any information. 
\begin{figure}[htbp]
	\includegraphics[width=0.45\textwidth]{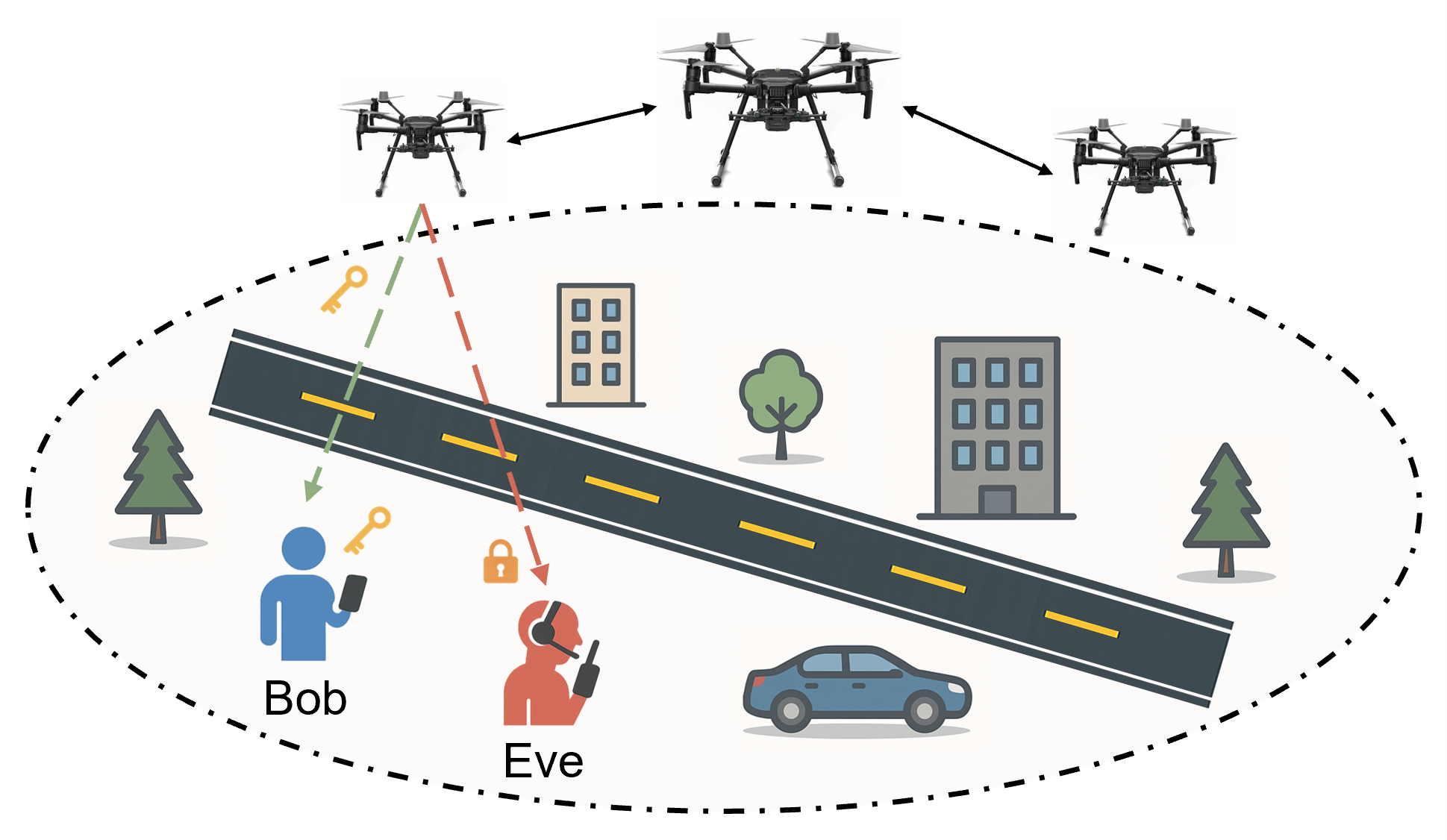}
	\centering
	\caption{ System model of UAV secure communications.}
	\label{fig1}
\end{figure}

\par Let $h_{ij}(t),~ i,j \in \{A,B\}$ denote the CSI from node $i$ to node $j$ at time $t$. Alice and Bob operate in a half-duplex fashion and alternately transmit pilot signals to each other to sample the CSI. The corresponding channel estimation are given by
\begin{equation}
	\hat{h}_A(t_1) = h_{BA}(t_1 - \tau_{BA}) + z_A(t_1),
	\label{eq:hA_est}
\end{equation}
\begin{equation}
	\hat{h}_B(t_2) = h_{AB}(t_2 - \tau_{AB}) + z_B(t_2),
	\label{eq:hB_est}
\end{equation}
where $\hat{h}_A$ and $\hat{h}_B$ denote the CSI observed by Alice and Bob, respectively. ${t}_1$ and ${t}_2$ are the time instants when two consecutive pilot signals arrive at Alice and Bob. $\tau_{ij},~ i,j \in \{A,B\}$ denotes the propagation delay from node $i$ to node $j$, with $\tau_{AB} \approx \tau_{BA}$. $t_1 - \tau_{BA}$ and $t_2 - \tau_{AB}$ represent the time instants when Bob and Alice transmit their pilot signals, respectively. The terms $z_A$ and $z_B$ denote the receiver noise at Alice and Bob, which are mutually independent and modeled as:
\begin{equation}
	z_i \sim \mathcal{CN}(0,\sigma_i^{2}), \quad i \in \{A,B\}, 
\end{equation} 
where $\sigma_i^{2}$ represents the noise power.
\par When the delay $\tau = |t_2 - t_1|$ is smaller than the channel coherence time, the underlying propagation channel between Alice and Bob can be regarded as approximately constant, and ideal channel reciprocity would imply
\begin{equation}
	h_{AB}(t_1 - \tau_{BA}) = h_{BA}(t_2 - \tau_{AB}).
\end{equation}
\par However, in UAV communications, the high mobility and interference may impair channel reciprocity. The estimated CSI at Alice and Bob are modeled as
\begin{equation}
	\hat{h}_A(m) = {h}(m) + w_A(m),
\end{equation}
\begin{equation}
	\hat{h}_B(m) = {h}(m) + w_B(m),
\end{equation}
where $h(m)$ denotes the common reciprocal channel coefficient, $w_A(m)$ and $w_B(m)$ are zero-mean Gaussian random variables that account for the channel estimation errors. The legitimate users repeat the above channel estimation  until they obtain the sequences of length $M$. In this paper, we assume that the channel follow the Rayleigh fading distribution.

\par For an eavesdropper, the main channel and the eavesdropping channel exhibit a certain degree of correlation due to the shared propagation environment and similar line-of-sight components \cite{kojima2024random}. To investigate the impact of this correlation on the security of UAV communications, we evaluate the proposed scheme under different channel correlation coefficients. The eavesdropping channel is modeled as follows:
\begin{equation}
	\hat{h}_E = \left( \rho \hat{h}_B + \sqrt{1-\rho^{2}}\, v \right) + n_{e},
\end{equation}
where $\rho$ denotes the correlation coefficient between the eavesdropping channel and the main channel, with $\rho \in [0,1]$; $v \sim \mathcal{CN}(0,1)$ is a Gaussian random variable with zero mean and unit variance; and $n_e$ represents the equivalent receiver noise at the eavesdropper.

\section{Secret Key Generation Based on Channel Codeword Reconstruction}
\par In this section, we first introduce the motivation of our design by showing how reconstructing channel codeword can mitigate the impact of channel reciprocity degradation, and then present a specific codeword reconstruction scheme for generating more consistent secret keys.

\subsection{Design Motivation}\label{AA}
\par Current PLKG schemes most rely on quantization of reciprocal channel characteristics to generate keys. However, when channel reciprocity degrades, the two legitimate parties obtain different observations of the channel. Directly applying existing schemes in this case often leads to inconsistent quantization results, which will force the discard of useful channel information, and further reduce the KGR. Quantization errors are usually eliminated through multiple rounds of key reconciliation, but the information exchanges not only increase system overhead, they also introduce potential security vulnerabilities. 

\par To address the above issues, the proposed scheme introduces the concept of channel codeword reconstruction. Specifically, the channel observed by Bob can be modeled as Alice's channel superimposed with noise. By reconstructing it into a codeword structure, we can leverage the powerful capabilities of ECC to mitigate the impact of noise and other interference. Furthermore, by applying tail truncation, the unreliable keys that are most affected by reciprocity degradation are discarded, while the reliable positions are preserved. In this way, our proposed scheme effectively mitigates the impact of channel reciprocity degradation and yields more consistent keys.
\subsection{Proposed scheme}

\par The proposed scheme includes five steps: channel probing, initial codeword generation, permutation-table construction, channel codeword decoding and adaptive tail truncation. The overall flowchart of the proposed scheme is illustrated in Fig.~2. In this paper, we consider a code with rate $R=k/n$, meaning that $k$ input bits are mapped to $n$ output bits. We denote the encoder by $Enc(\cdot)$ and the decoder by $Dec(\cdot)$.
\begin{figure}[htbp]
	\includegraphics[width=0.45\textwidth]{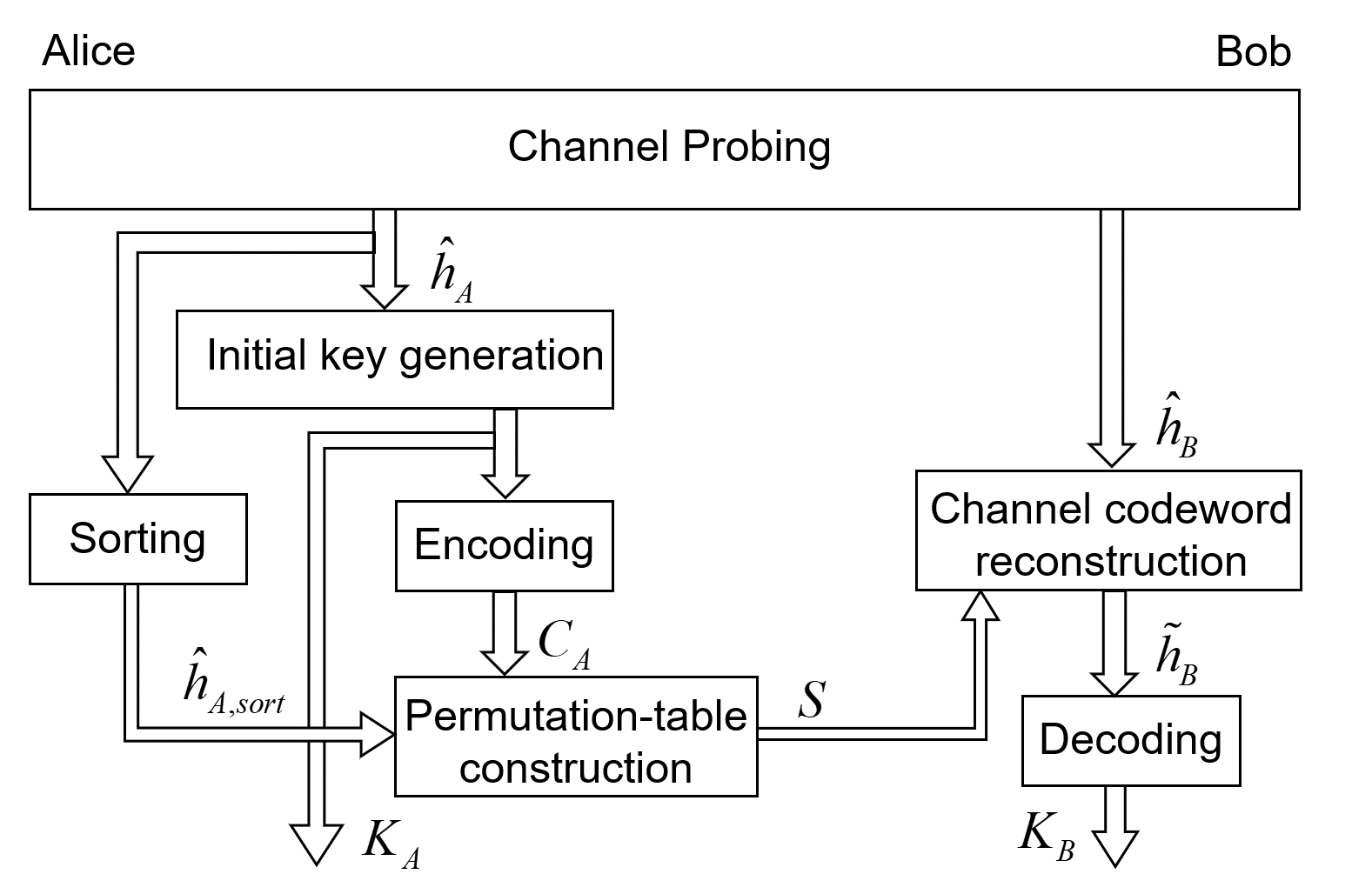}
	\centering
	\caption{ Flowchart of the proposed key generation scheme.}
	\label{fig1}
\end{figure}

\subsubsection{Channel Probing}
\par Alice and Bob alternately transmit pilot signals to each other. Upon receiving a pilot signal, each node estimates the CSI. As multiple rounds of exchanges, Alice and Bob obtain $M$ observations of the channel, denoted by $\hat{h}_A$ and $\hat{h}_B$. 
\subsubsection{Initial Codeword Generation}
\par The initial codeword is constructed from a locally generated binary key at Alice. Specifically, Alice extracts an initial key from her channel estimation $\hat{h}_A$ as
\begin{equation}
	K_{\text{A}}[i] =
	\begin{cases}
		0, & \hat{h}_{{A}}[i] > 0,\\
		1, & \hat{h}_{{A}}[i] \le 0,
	\end{cases}
	\quad i = 1,2,\ldots,\left\lfloor \dfrac{kM}{n} \right\rfloor,
\end{equation}
or equivalently from an uniform binary source. The sequence satisfies $K_{\text{A}} \in \{0,1\}^{\left\lfloor \frac{kM}{n} \right\rfloor}$ and is then encoded to obtain the initial codeword 
\begin{equation}
	C_{A} = Enc(K_{A}).
\end{equation}

\subsubsection{Permutation-Table Construction}
\par Alice constructs the permutation-table $S$ based on the initial codeword $C_{A}$ and the channel estimation $\hat{h}_{{A}}$. Firstly, sort $\hat{h}_{A}$ in ascending order to obtain $[\hat{h}_{A,\mathrm{sort}}, {T}_{\mathrm{sort}}]=\mathrm{sort}(\hat{h}_A)$, where $\hat{h}_{A,\mathrm{sort}}$ denotes the sorted channel estimation and ${T}_{\mathrm{sort}}$ records the indices of the elements of $\hat{h}_{A,\mathrm{sort}}$ in the original vector $\hat{h}_A$. 
\par Following the sorting process, the permutation-table is constructed according to Algorithm 1. The resulting Table $S$ is then transmitted to the receiver over the wireless channel.
\begin{algorithm}[!h]
	\caption{Permutation-Table Construction Algorithm}
	\label{alg:codeword-map}
	\begin{algorithmic}[1]
		\State \textbf{Input:} the codeword $C_A$; the indices ${T}_{\mathrm{sort}}$
		\State \textbf{Output:} the permutation-table $S$
		\State Initialize pointers: $p_0 \gets 1$, $p_1 \gets M$
		\State Initialize iterator: $i \gets 1$
		\Repeat
		\If{$C_A[i] = 0$}
		\State $S[i] \gets {T}_{\mathrm{sort}}[p_0]$; \quad $p_0 \gets p_0 + 1$;
		\Else
		\State $S[i] \gets {T}_{\mathrm{sort}}[p_1]$; \quad $p_1 \gets p_1 - 1$;
		\EndIf
		\State $i \gets i + 1$;
		\Until{$i = \frac{n}{k}\lfloor \frac{kM}{n} \rfloor$}
		\State \textbf{return} $S$
	\end{algorithmic}
\end{algorithm}
\subsubsection{Channel Codeword Decoding}
\par Upon receiving the permutation-table $S$, Bob reorders the estimated channel $\hat{h}_B$ as
\begin{equation}
	\tilde{h}_B[i] = \hat{h}_B[S[i]].
\end{equation}
\par This operation produces a randomized channel sequence with a codeword-like structure. Bob generates the secret key by decoding this sequence as:
\begin{equation}
	K_B=Dec(\tilde{h}_B).
\end{equation}
\par We employ soft-decision decoding for the decoder $Dec(\cdot)$. Unlike hard-decision,  $\tilde{h}_B$ conveys soft reliability information. Larger entries in $\tilde{h}_B$ correspond to a higher confidence level for the binary symbol 1', while smaller entries are considered closer to `0'.

\subsubsection{Adaptive Tail Truncation}
\par Alice and Bob retain the high-consistency keys through tail truncation.
Specifically, Alice determines the optimal key length $L_{\text{opt}}$, based on the observed channel quality and simulation results. The final keys ${K}'_{A}$ and ${K}'_{B}$ are derived by truncating the initial sequences to the specified length
\begin{equation} {K}'_{A} = {K}_{A}[1 : L_{\text{opt}}], \quad {K}'_{B} = {K}_{B}[1 : L_{\text{opt}}]. \end{equation}

\subsection{Selection of Error-Correcting Code}
\par The main existing ECC include polar code, LDPC code, and convolutional code, etc. Specifically, polar code exploit channel polarization for decoding, while LDPC code perform iterative decoding by exchanging and updating messages in each iteration. In the proposed scheme, the sorting process results in a channel distribution that strictly varies with the sequence index. This ordered non-uniformity fundamentally differs from the position-independence assumption that underlie Polar and LDPC code.
\par For convolutional code, the Viterbi decoding algorithm uses the distance between the received signal and each candidate codeword as the path metric, and selects the optimal path with the minimum distance. In this case, the bit with the poorest quality dominates the difference between candidate paths. Therefore, if the more reliably received signal are placed earlier, the decoding accuracy is higher than when they are placed near the end. This mechanism naturally matches the channel-sorting property of the proposed scheme, making convolutional code the most suitable choice.

\subsection{Theoretical Analysis}
\par In this section, we derive the probability distributions of the channels in ascending order and demonstrate that the proposed scheme achieves a polarization characteristic analogous to Polar codes. This characteristic effectively segregates reliable keys from unreliable ones, which provides the rationale for the proposed tail truncation strategy.

\par For general order statistics, let $F(x)$ and $f(x)$ denote the cumulative distribution function (CDF) and probability density function (PDF) of the population $X$, respectively. Let $X_1, X_2, \cdots, X_M$ represent $M$ independent and identically distributed samples drawn from this population, sorted in ascending order as
\begin{equation}
	X_{(1)} \le X_{(2)} \le \cdots \le X_{(M)}.
\end{equation}
\par Defining $X_{(1)}, X_{(2)}, \cdots, X_{(M)}$ as the order statistics, the PDF of the $m$-th order statistic $X_{(m)}$ can be derived as
\begin{equation}
	f_m(x) = \frac{M!}{(m-1)!(M-m)!} [F(x)]^{m-1} [1-F(x)]^{M-m} f(x).
\end{equation}
\par For a random variable $h$ following a Gaussian distribution $h \sim \mathcal{N}(\mu, \sigma^2)$, the PDF is given by 
\begin{equation}
	f_H(h) = \frac{1}{\sqrt{2\pi}\sigma} \exp\left(-\frac{(h-\mu)^2}{2\sigma^2}\right),
\end{equation}
while its CDF denoted as $F_H(h) = \int_{-\infty}^h f_H(x) \mathrm{d}x$.
\par Since a closed-form solution is mathematically intractable using the general formula derived above, we employ numerical computation. Consequently, the PDF of the sorted statistic $h_{(m)}$ is evaluated as follows:
\begin{equation}
	\small
	\begin{aligned}
		f(h) &= \frac{M!}{(m-1)!(M-m)!} [F_H(h)]^{m-1} [1-F_H(h)]^{M-m} f_H(h) \\
		&= \frac{(\prod_{i=1}^{m-1} i) \times m \times \prod_{i=m+1}^{M} i}{(\prod_{i=1}^{m-1} i) \times 1 \times \prod_{i=1}^{M-m} i} \\
		& \quad \times [F_H(h)]^{m-1} [1-F_H(h)]^{M-m} f_H(h). \\
		&= [F_H(h)]^{m-1} \times m f_H(h) \times \prod_{i=1}^{M-m} \left[ \frac{m+i}{i} (1-F_H(h)) \right]
	\end{aligned}
	\label{eq:gaussian_order_pdf_derivation}
\end{equation}
\par To prevent numerical overflow caused by large factorials in the theoretical form, the third equality presents a implementation using factorial decomposition. Fig.~3 illustrates the distribution of order statistics with a channel sample size of $M=100$.
\begin{figure}[htbp]
	\includegraphics[width=0.4\textwidth]{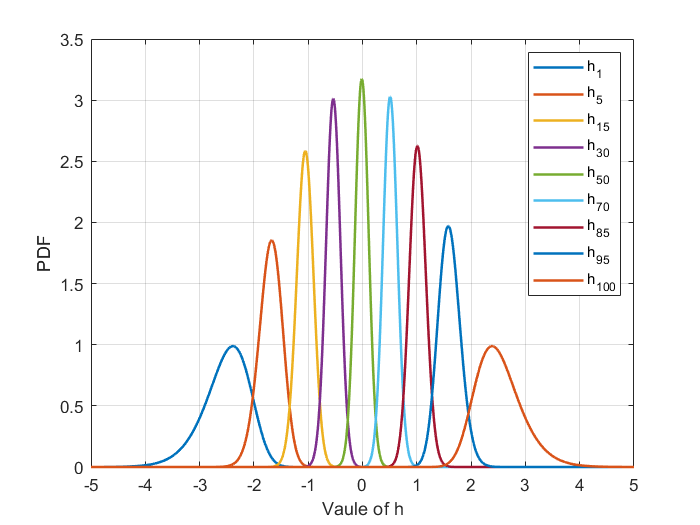}
	\centering
	\caption{The PDF of the order statistic $h_{(m)}$.}
\end{figure}
\par As observed from the Fig.~3, the extremities samples of the sorted sequence correspond strongly to '0' and '1', which means these bits virtually error-free. The proposed scheme achieve a polarization characteristic analogous to Polar codes, effectively segregating reliable keys from unreliable ones. This separation provides the theoretical basis for the tail truncation strategy, which is designed to preserve the extremities samples. 
\par In our proposed scheme, after the permutation table construction, Alice uses Table $S$ to reorder the original channel into $\widetilde{h}_A$. After Bob applies the permutation to his observations, the $i$-th element of his reordered channel is given by
\begin{equation}
	\widetilde{h}_B[i] = \widetilde{h}_A[i] - w_A + w_B,
\end{equation}
where the distributions of $w_A$ and $w_B$ are independent of the sorted position and continue to follow their original noise distributions. Given $\widetilde{h}_A[i]$, the conditional PDF of $\widetilde{h}_B[i]$ is given by
\begin{equation}
	f_{\widetilde{H}_B[i]}(\widetilde{h}_B[i] \mid \widetilde{h}_A[i]) = \frac{1}{\sqrt{2\pi(\sigma_A^2 + \sigma_B^2)}} e^{-\frac{(\widetilde{h}_B[i] - \widetilde{h}_A[i])^2}{2(\sigma_A^2 + \sigma_B^2)}}.
\end{equation}
\par This indicates that the noise power of $\widetilde{h}_B[i]$ is the superposition of the noise powers from both the transmitter and the receiver. When the noise powers are symmetric, i.e., $\sigma_A = \sigma_B = \sigma$, the equation simplifies to
\begin{equation}
	f_{\widetilde{H}_B[i]}(\widetilde{h}_B[i] \mid \widetilde{h}_A[i]) = \frac{1}{\sqrt{4\pi\sigma^2}} e^{-\frac{(\widetilde{h}_B[i] - \widetilde{h}_A[i])^2}{4\sigma^2}}
	\label{eq:conditional_pdf_symmetric}
\end{equation}
\par It is crucial to note that this noise power is independent of the position $i$. This implies that for CSI at certain sorted positions, the noise may dominate. Consequently, the strategy of tail truncation is highly effective for the proposed scheme, as it discards unreliable keys and enables highly consistent key generation even under severe channel estimation errors.

\section{SIMULATION RESULTS AND DISCUSSIONS}
\par In this section, we evaluate the performance of the proposed PLKG scheme based on channel codeword reconstruction. Our evaluation focuses on two primary metrics: KDR and the Number of Generated Keys.

\par {1) Key Disagreement Rate (KDR):} The KDR quantifies the discrepancy between the initial keys ${K}_A$ and ${K}_B$, which is formally defined as:
\begin{equation}
	\text{KDR} = \frac{1}{L}\sum_{i=1}^{L} \left( K_A[i] \oplus K_B[i] \right),
	\label{eq:KDR}
\end{equation}
where $\oplus$ represents the exclusive OR (XOR) operation.

\par {2) Number of Generated Keys:} The number of generated keys represents the total effective key length retained after tail truncation. In the proposed scheme, the polarization characteristic facilitates a trade-off between the KDR and the number of keys, which enables the selection of an appropriate truncation ratio tailored to specific requirements.

\subsection{Impact of Tail Truncation on Key Generation Performance }
\par To the best of our knowledge, channel-reconstruction-based schemes have not been reported in the existing literature. For comparison, we consider an uncoded scheme, a repetition-code scheme, and the convolution-code scheme, and evaluate the performance of these three methods under a rate of 1/3 and an channel sample size of 600. In the following analysis, we evaluate the KDR performance between legitimate users under varying truncation ratios.
\begin{figure}[htbp]
	\includegraphics[width=0.4\textwidth]{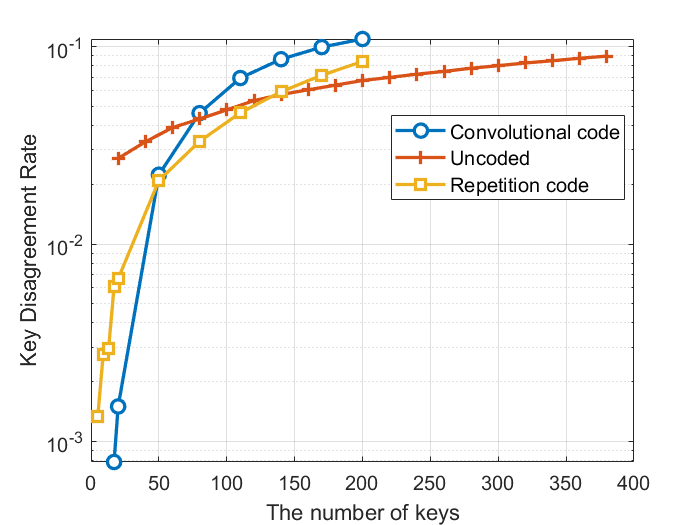}
	\centering
	\caption{KDR performance under different tail truncation ratios.}
	\label{fig1}
\end{figure}
\par The Fig.~4 illustrates the relationship between the number of key and the KDR at an signal-to-noise ratio (SNR) of 0 dB. Under low-SNR condition, the uncoded scheme achieves the maximum KDR, followed by the repetition-code scheme. The convolutional-code scheme demonstrates the superior performance by yielding the lowest KDR.
\par To further evaluate the proposed scheme, we analyze the number of generated keys at different tail truncation ratios where KDR is below $10^{-3}$, as shown in Fig.~5. At low-SNR condition, convolutional code can generate more consistent keys. Repetition code exhibit similar properties to convolutional code, but its performance is consistently inferior.
\begin{figure}[htbp]
	\includegraphics[width=0.4\textwidth]{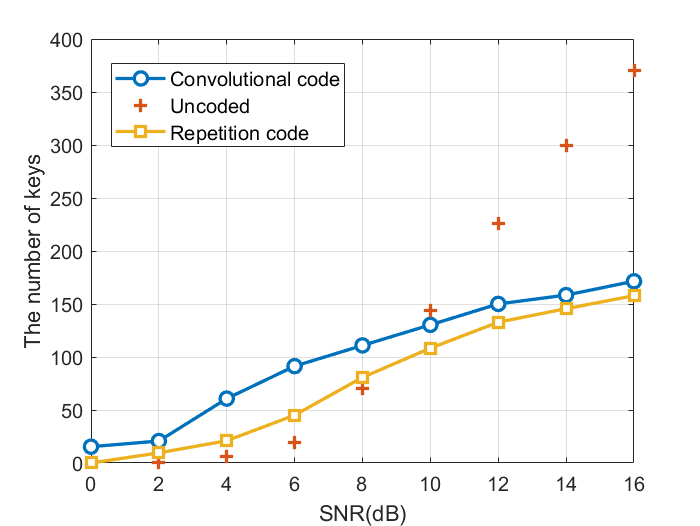}
	\centering
	\caption{The number of generated keys under different tail truncation ratios.}
	\label{fig1}
\end{figure}
\par The above simulation results demonstrate that, under low-SNR conditions, the proposed scheme exhibits superior robustness, generating more consistency keys than the baseline.
\subsection{KDR Performance versus SNR for Different Schemes} 
\label{subsec:ss}
\par As baseline schemes for comparison, we adopt an uncoded scheme, a single-threshold quantization scheme, and the level-crossing algorithm (LCA) with with a threshold factor of 0.4. Fig.~6 compares the KDR performance across varying SNRs, where the proposed scheme operates at a coding rate of 1/2. To ensure a fair comparison, the retained length of the proposed scheme is configured to be identical to that of the LCA scheme.
\begin{figure}[htbp]
	\includegraphics[width=0.4\textwidth]{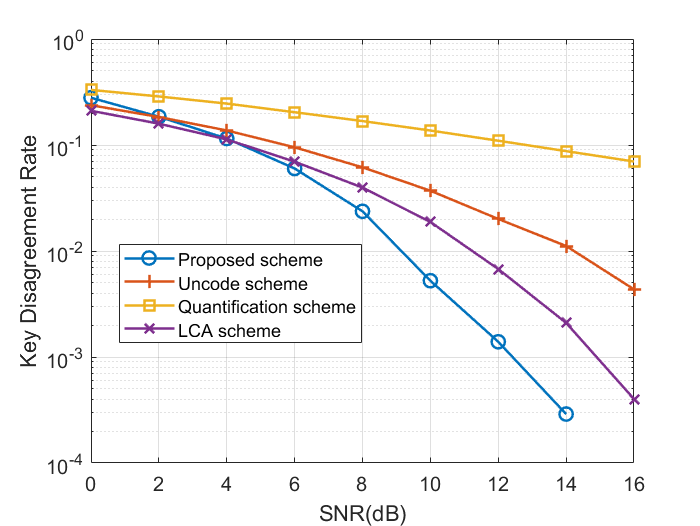}
	\centering
	\caption{Alice–Bob KDR under different SNRs.}
	\label{fig1}
\end{figure}

\par The performance gap between the proposed scheme and the baseline benchmarks widens as the SNR increases. The KDR of the proposed scheme decreases much more rapidly, indicating superior error-correction capability. To achieve the same KDR, the proposed scheme provides a performance gain of approximately 2 dB over the LCA scheme. These results confirms that the proposed scheme significantly enhances key reliability under equivalent channel conditions.

\subsection{Security performance}
\par To comprehensively evaluate the security performance, we assume a powerful eavesdropper model in which Eve can intercept the permutation-table transmitted by Alice. We then analyze the KDR between Alice and Eve under different channel correlation coefficients. We adopt the same set of baseline schemes and parameter configurations as detailed in Section \ref{subsec:ss}.
\begin{figure}[htbp]
	\includegraphics[width=0.4\textwidth]{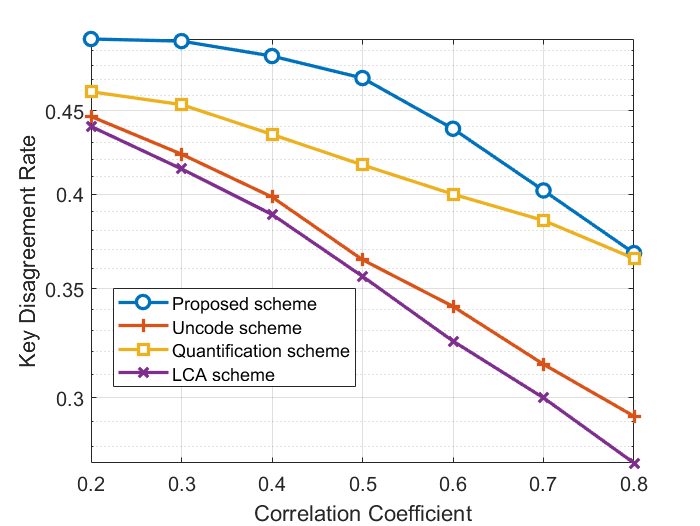}
	\centering
	\caption{Alice–Eve KDR under different correlation coefficients.}
	\label{fig1}
\end{figure}
\par For low-to-moderate correlation coefficients in the range from 0.2–0.6, the proposed scheme achieves an Alice–Eve KDR above 0.45, indicating that Eve can hardly obtain any useful key information. As the channel correlation coefficient further increases, the security advantage of the proposed scheme becomes more pronounced, with its KDR being up to approximately 36\% higher than that of the LCA scheme. Consequently, the proposed scheme offers the best security performance among all the baseline schemes.

\section{Conclusion}
\par In this paper, we proposed a novel PLKG scheme based on channel codeword reconstruction for UAV communications, which exhibits a polarization property that enables the separation of reliable and unreliable keys. Simulation results show that, compared with repetition-code and uncoded schemes, the proposed approach achieves a significantly lower KDR while generating more consistent keys. We also provides a higher KDR at the eavesdropper than LCA scheme, which enhances resistance to eavesdropping. Considering the inherent trade-off between the KGR and KDR, an appropriate tail truncation ratio should be selected according to the requirements of specific applications.

\bibliographystyle{IEEEtran}
\bibliography{References} 

\begin{thebibliography}{10}
\providecommand{\url}[1]{#1}
\csname url@samestyle\endcsname
\providecommand{\newblock}{\relax}
\providecommand{\bibinfo}[2]{#2}
\providecommand{\BIBentrySTDinterwordspacing}{\spaceskip=0pt\relax}
\providecommand{\BIBentryALTinterwordstretchfactor}{4}
\providecommand{\BIBentryALTinterwordspacing}{\spaceskip=\fontdimen2\font plus
\BIBentryALTinterwordstretchfactor\fontdimen3\font minus
  \fontdimen4\font\relax}
\providecommand{\BIBforeignlanguage}[2]{{%
\expandafter\ifx\csname l@#1\endcsname\relax
\typeout{** WARNING: IEEEtran.bst: No hyphenation pattern has been}%
\typeout{** loaded for the language `#1'. Using the pattern for}%
\typeout{** the default language instead.}%
\else
\language=\csname l@#1\endcsname
\fi
#2}}
\providecommand{\BIBdecl}{\relax}
\BIBdecl

\bibitem{yao2021resource}
Z.~Yao, W.~Cheng, W.~Zhang, and H.~Zhang, ``Resource allocation for
  5g-uav-based emergency wireless communications,'' \emph{IEEE Journal on
  Selected Areas in Communications}, vol.~39, no.~11, pp. 3395--3410, 2021.

\bibitem{zheng2025optimal}
Y.~Zheng, L.~Li, W.~Lin, W.~Liang, Q.~Du, and Z.~Han, ``Optimal transport
  framework for isac in low-altitude networks: Joint resource allocation for
  cooperative communication and non-cooperative localization,'' \emph{IEEE
  Transactions on Communications}, vol.~74, pp. 1984--2000, 2025.

\bibitem{ceviz2024survey}
O.~Ceviz, S.~Sen, and P.~Sadioglu, ``A survey of security in uavs and fanets:
  Issues, threats, analysis of attacks, and solutions,'' \emph{IEEE
  Communications Surveys \& Tutorials}, 2024.

\bibitem{sen2023cryptography}
J.~Sen, ``Cryptography and key management schemes for wireless sensor
  networks,'' \emph{arXiv preprint arXiv:2307.00872}, 2023.

\bibitem{sun2018review}
L.~Sun and Q.~Du, ``A review of physical layer security techniques for internet
  of things: Challenges and solutions,'' \emph{Entropy}, vol.~20, no.~10, p.
  730, 2018.

\bibitem{sun2017physical}
------, ``Physical layer security with its applications in 5g networks: A
  review,'' \emph{China communications}, vol.~14, no.~12, pp. 1--14, 2017.

\bibitem{xiao2023secure}
Y.~Xiao, Q.~Du, W.~Cheng, and N.~Lu, ``Secure communication guarantees for
  diverse extended-reality applications: A unified statistical security
  model,'' \emph{IEEE Journal of Selected Topics in Signal Processing},
  vol.~17, no.~5, pp. 1007--1021, 2023.

\bibitem{xiao2024statistical}
Y.~Xiao, Q.~Du, and G.~K. Karagiannidis, ``Statistical age of information: A
  risk-aware metric and its applications in status updates,'' \emph{IEEE
  Transactions on Wireless Communications}, vol.~24, no.~3, pp. 2325--2340,
  2024.

\bibitem{xiao2023adaptive}
Y.~Xiao, Q.~Du, W.~Cheng, and W.~Zhang, ``Adaptive sampling and transmission
  for minimizing age of information in metaverse,'' \emph{IEEE Journal on
  Selected Areas in Communications}, vol.~42, no.~3, pp. 588--602, 2023.

\bibitem{gao2024ris}
N.~Gao, Y.~Yao, S.~Jin, C.~Li, and M.~Matthaiou, ``Ris-assisted simultaneous
  transmission and secret key generation: an icas paradigm,'' in \emph{IEEE
  INFOCOM 2024-IEEE Conference on Computer Communications Workshops (INFOCOM
  WKSHPS)}.\hskip 1em plus 0.5em minus 0.4em\relax IEEE, 2024, pp. 1--6.

\bibitem{li2019hybrid}
G.~Li, Z.~Zhang, Y.~Yu, and A.~Hu, ``A hybrid information reconciliation method
  for physical layer key generation,'' \emph{Entropy}, vol.~21, no.~7, p. 688,
  2019.

\bibitem{liu2019secret}
B.~Liu, A.~Hu, and G.~Li, ``Secret key generation scheme based on the channel
  covariance matrix eigenvalues in fdd systems,'' \emph{IEEE Communications
  Letters}, vol.~23, no.~9, pp. 1493--1496, 2019.

\bibitem{torshizi2024exploiting}
E.~O. Torshizi and W.~Henkel, ``Exploiting fdd channel reciprocity for physical
  layer secret key generation in iot networks,'' \emph{IEEE Communications
  Letters}, vol.~28, no.~6, pp. 1268--1272, 2024.

\bibitem{he2022deep}
H.~He, Y.~Chen, X.~Huang, M.~Xing, Y.~Li, B.~Xing, and L.~Chen, ``Deep
  learning-based channel reciprocity learning for physical layer secret key
  generation,'' \emph{Security and Communication Networks}, vol. 2022, no.~1,
  p. 1844345, 2022.

\bibitem{kojima2024random}
S.~Kojima and S.~Sugiura, ``Random pilot activation and interpolated channel
  estimation for physical-layer secret key generation in correlated
  eavesdropping channel,'' \emph{IEEE Transactions on Vehicular Technology},
  vol.~73, no.~9, pp. 12\,978--12\,990, 2024.

\end{thebibliography}

\end{document}